\documentclass{pasa}%

\usepackage{graphicx}

\title[Calibration of EFOSC2 broadband linear imaging polarimetry]{Calibration of EFOSC2 broadband linear imaging polarimetry}
\author[K. Wiersema et al.]{K. Wiersema$^{1,2}$ \thanks{E-mail:
K.Wiersema@warwick.ac.uk}, A. B. Higgins$^{1}$, S. Covino$^{3}$, R. L. C. Starling$^{1}$
\affil{$^1$ Department of Physics \& Astronomy and Leicester Institute of Space \& Earth Observation, University of Leicester, University Road, Leicester LE1 7RH, United Kingdom}%
\affil{$^2$ Department of Physics, University of Warwick, Coventry, CV4 7AL, UK}
\affil{$^3$ INAF/Brera Astronomical Observatory, via Bianchi 46, I-23807, Merate (LC), Italy}
}%

\jid{PASA}
\doi{10.1017/pas.\the\year.xxx}
\jyear{\the\year}

\usepackage{aas_macros}
\usepackage{hyperref} 
\hypersetup{colorlinks,citecolor=blue,linkcolor=blue,urlcolor=blue}

\begin{document}

\begin{frontmatter}
\maketitle

\begin{abstract}
EFOSC2 (the European Southern Observatory Faint Object Spectrograph and Camera v2) is one of the workhorse instruments 
on ESO's New Technology Telescope (NTT), and is one of the most popular instruments at La Silla observatory.
It is mounted at a Nasmyth focus, and therefore exhibits strong, wavelength and pointing-direction dependent instrumental polarisation.
In this document we describe our efforts to calibrate the broadband imaging polarimetry mode, and provide a calibration for broadband $B, V, R$ filters
to a level that satisfies most use cases (i.e. polarimetric calibration uncertainty $\sim 0.1\%$). We make our calibration codes public. 
This calibration effort can be used to enhance the yield of future polarimetric programmes with EFOSC2, by allowing good calibration with
a greatly reduced number of standard star observations. Similarly, our calibration model can be combined with archival calibration observations 
to post-process data taken in past years, to form a EFOSC2 legacy archive with substantial scientific potential.
\end{abstract}

\begin{keywords}
instrumentation: polarimeters -- techniques: polarimetric 
\end{keywords}
\end{frontmatter}

\section{INTRODUCTION }
\label{sec:intro}
The European Southern Observatory Faint Object Spectrograph and Camera v2 (EFOSC2) is a highly versatile, focal-reducer based, instrument, capable of
both spectroscopy and imaging at high efficiency levels. Its design is based on EFOSC (\citealt{Buzzoni}), which was developed for 
the ESO 3.6m telescope at La Silla (Chile). After EFOSC was decommissioned, EFOSC2 spent a few semesters mounted at the 3.6m; then was offered at the New Technology Telescope (NTT; with a 3.58 meter primary mirror) from April 2008 until now.  
 Both imaging and spectroscopy are available for EFOSC2, and in both cases it is possible to use wave plates and a Wollaston element to
perform polarimetry: imaging polarimetry and spectropolarimetry (for detail see the EFOSC2 user manual, \citealt{usermanual}).  Over the many semesters that EFOSC2 has been available, a formidable
archive of data has been built up, which includes several polarimetric observing programmes. Unfortunately, a detailed calibration of the instrumental
linear polarisation behaviour of EFOSC2 has so far not been available in the literature, which may discourage future users and affect the efficiency of their programmes. 
Because of its mounting on a Nasmyth platform, EFOSC2 has high levels of wavelength dependent instrumental polarisation which depend on the direction that the telescope is 
pointing (e.g. \citealt{Tinbergen}). This can be calibrated accurately, using a large set of standard star observations. The resulting polarisation model can then be used as a starting point
for future, and past, observations, which can use a much smaller set of standard star observations to tweak the model to their observing epoch, saving 
valuable observing time and increasing the accuracy of calibration. In this paper we aim to provide such a calibration of EFOSC2 broadband linear imaging 
polarimetry, and make the resulting Python 3\,\footnote{\tt www.python.org} codes publicly available\footnote{The codes used for this analysis are available from the authors, or alternatively available at {\tt github.com/abh13/EFOSC2$\_$Scripts}}. 

\section{Instrument}
\begin{figure}
\centerline{\includegraphics[width=8cm]{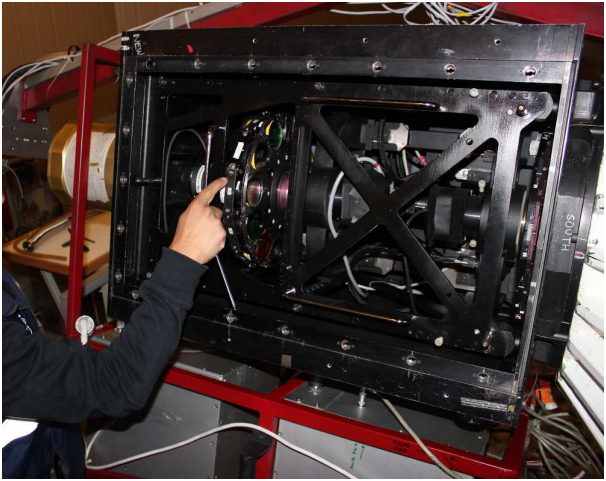}}
\caption{The insides of EFOSC2, mounted on the Nasmyth focus of the NTT. Light enters from the right, the camera is on the left.
In between, the two wheels containing filters and grisms can be seen, the finger points at the Wollaston element that was used 
for the EFOSC2 imaging polarimetry in this paper. }
\label{fig:inside}
\end{figure}

A thorough description of the EFOSC2 instrument can be found in the EFOSC2 user manual (\citealt{usermanual}), here we 
summarize some of the relevant properties. To obtain polarimetric measurements, EFOSC2 has
super-achromatic half- and quarter wave plates, and Wollaston elements with 20 and 10 arcsecond 
beam separations. The Wollaston element splits the light into so-called ordinary and extraordinary beams ($o$ and $e$ beams, hereafter), with orthogonal polarisation, which are recorded simultaneously by the camera. In this work, we use the Wollaston element with 20 arcsecond separation (``Woll\_Prism20'' in the ESO headers). A Wollaston mask, made of alternating open and closed strips, is used to avoid the two images (of the two beams) overlapping; it is mounted in the slit wheel. For imaging polarimetry, the Wollaston element is mounted in the grism wheel,  see the EFOSC2 user manual (\citealt{usermanual},  their figure 1) for a schematic of the instrument parts, and Figure \ref{fig:inside} for a photo of the wheels inside EFOSC2 to give the relevant scales. 
In this work, the Wollaston element and the mask (``WollMask='') are mounted such that the strips are
parallel to the CCD x-axis (East-West direction: the orientation of the EFOSC2 data is such that North is up, and East to the right). We define the $o$ and $e$ beam images to be the top and bottom images, respectively. 
We used a $2\times2$ binning of the CCD, which results in a effective pixel scale of 0.24 arcsecond per pixel, and a field of view of $\sim 4.1\times 4.1$ arcminutes, though the use of the Wollaston mask means that only half of the field of view (that which falls in the open strips of the mask) is recorded in each exposure.

\begin{table*}
 \centering
  \caption{Standard stars observed in the 2016 observing run. All standards were observed in B, V and R bands. Object names in italic font identify the polarised standards, the other objects are zero polarisation standard stars. The adopted Stokes $q$ and $u$ values for the polarised standards (second and third column of this table) are taken from \cite{Fossati}. We also list the uncertainties given by \cite{Fossati} on their $q,u$ measurements for these stars. For the unpolarised standards we adopt $q = u = 0$ for all three bands; see Sections \ref{sec:unpolarisedstds} and \ref{sec:polarisedstds} for a discussion.
  \label{table:obslog1}}
 \begin{tabular}{llllllll} 
  \hline
Object                          & Adopted $q$ & Adopted $u$  &   Mag.     &  Exp. time  \\ 
                        & ($\times 100\%$) & ($\times 100\%$) &     (V)      &  (s)       \\
 \hline  
{\it BD$-$12 5133}          &$B: 1.87\pm0.04$  & $-3.95\pm0.05$ &    10.4    &   1        \\
                            &$V: 1.75\pm0.04$  & $-4.00\pm0.04$  &        &                \\
                            &$R: 1.63\pm0.02$  & $-3.68\pm0.02$  &        &                \\
{\it Hilt 652 (CD$-$28 13479)} &$B: 5.70\pm0.01$ &  $-0.11\pm0.03$  &    10.8     &  1, 2  \\
                               &$V: 6.24\pm0.03$ &  $-0.18\pm0.04$  &        &             \\
                               &$R: 6.07\pm0.02$ &  $-0.18\pm0.04$  &        &             \\
{\it Vela 1 95 (Ve 6$-$23)} &$B: 7.12\pm0.05$  & $-1.66\pm0.03$     &    12.1     &   2    \\
                            &$V: 7.91\pm0.05$  & $-2.38\pm0.06$     &        &             \\
                            &$R: 7.56\pm0.06$  & $-2.32\pm0.03$    &        &             \\
\hline
WD 1344$+$106               & &                  &    15.1     &    20         \\     
WD 1615$-$154                & &                &    13.4     &    4           \\
WD 1620$-$391                & &                &    11.0     &    2           \\      
WD 2039$-$202                & &                &    12.4     &    2           \\     
WD 2359$-$434                & &                &    13.0     &    3           \\
 \hline
\end{tabular}
\normalsize
\end{table*}

\begin{table*}
 \centering
  \caption{ Full log of unpolarised standard star observations used in this document. $q,u$ values are the measured instrumental values in the EFOSC2 
  coordinate system. ``mid'' denotes the value at the middle of the polarimetric sequence of four exposures.
 \label{table:unpol}}
 \begin{tabular}{llllllllll} 
  \hline
Object                                   & Filter   &   Obs. date      & Parallactic angle & $q$                        & $q$ error                & $u$                        & $u$ error \\ 
                                               &            &   (mid, MJD)    & (mid; degrees)    &  ($\times 100\%$)   &  ($\times 100\%$)   & ($\times 100\%$)    & ($\times 100\%$)   \\
 \hline  
WD 1344$+$106                 &     $V$       &    57558.9832     &  -160.01        &  -2.40    &     0.17         & 2.66  &    0.13             \\   
                                            &     $B$       &    57558.9859      &  -161.19        &   -2.68   &      0.26        & 2.10  &   0.20          \\      
                                            &     $R$       &   57558.9886      &   -162.39        &  -1.75    &      0.14        & 3.39  &   0.11          \\      
WD 1615$-$154                 &     $V$         &  57560.0979    &  -146.41            &  -3.07    &   0.15           & 0.96  &  0.12             \\
                                           &     $B$         &  57560.0999    &   -148.16           &  -2.95    &   0.16           & 0.59  &     0.12        \\
                                           &     $R$         &  57560.1019    &  -149.99            &  -3.29    &    0.15          &  1.94 &   0.12          \\
                                           &     $V$         &  57560.1614   &   142.14           &  3.08   &  0.16            & 1.19  &    0.13        \\
                                           &     $B$         &  57560.1634    &    140.68          &  2.40    &    0.16          & 1.22  & 0.13            \\
                                           &     $R$         &  57560.1653    &   139.30           & 3.82     &   0.16           & 0.39  & 0.13            \\
                                            &     $V$       &   57561.9884   &    -115.78          & -2.22     &   0.13           & -2.17  &  0.10           \\
                                            &     $B$       &   57561.9905   &    -115.88         &  -1.66    &     0.15         & -2.29  &   0.12          \\
                                            &     $R$       &   57561.9925   &     -115.99         &  -3.00    &  0.13            & -2.23  &  0.11           \\
                                            &     $V$      &  57562.1725    &       131.98        &  3.56    &   0.34           & -0.46  &  0.29           \\
                                            &     $B$      &   57562.1745    &      131.01         &  3.05    &   0.27           & 0.59  &  0.16           \\
                                            &     $R$      &   57562.1766   &        130.08       &  4.16    &  0.33            & -0.44  &  0.27           \\
WD 1620$-$391                  &     $V$        &  57559.0593   &      -73.86       &     1.91 &    0.07          & -2.60  &  0.05           \\    
                                            &     $B$        &  57559.0612    &      -73.09      &  2.04    &   0.07           & -2.08  &   0.06          \\
                                            &     $R$       &  57559.0631    &      -72.31       &  2.07    &   0.07           & -3.07  &  0.05           \\
                                            &     $V$        &  57559.1679    &    49.13         & -3.28     &   0.07           & -0.72  &  0.05           \\
                                            &     $B$        &  57559.1698    &    50.86         &  -2.60    &  0.07            & -1.32  &  0.05           \\
                                            &     $R$       &  57559.1716    &     52.47         &  -3.74    &   0.07           & -0.79  &    0.05         \\                                                                                                                                                                                              
                                            &    $V$        &  57560.0292   &       -83.04        &  1.05    &   0.10           & -3.07  &   0.07          \\      
                                            &     $B$       &  57560.0310   &     -82.51          &  1.36    &    0.10          & -2.77  &  0.08           \\      
                                            &     $R$       &  57560.0328   &     -81.98          &   0.72   &     0.09         & -3.60  &  0.07           \\      
WD 2039$-$202                  &    $V$        &   57559.2199  &  -113.55             &     -2.34 &    0.13          & -2.48  &   0.10          \\     
                                            &    $B$        &    57559.2218 &  -113.74             & -1.55     &   0.15           & -2.45  &  0.11           \\      
                                            &    $R$        &    57559.2237 &  -113.93             & -2.86     &    0.13          & -2.36  &     0.10        \\      
                                            &   $V$         &    57560.2688  &  -127.13           & -3.13     &     0.13         & -1.17  &    0.10         \\      
                                            &   $B$         &    57560.2707  &  -128.24           &  -2.63    &    0.15          & -1.38  &   0.11          \\      
                                            &   $R$         &   57560.2726   &  -129.42           &   -3.81   &     0.13         & -0.52  &    0.10         \\      
                                            &   $V$         &    57560.4149  &  112.67            & 2.35     &   0.14           &  -2.38 &   0.11          \\      
                                            &   $B$         &    57560.4168  &  112.56            &   2.66   &     0.16         &  -1.60 &   0.13          \\      
                                            &   $R$         &   57560.4187   &  112.46           &  2.24    &      0.15        & -3.05  &    0.11         \\      
WD 2359$-$434                  &    $V$        &   57559.3374   &  -80.87            &     1.25 &    0.15          & -3.07  &   0.12          \\
                                            &    $B$        &   57559.3393   &  -80.27            &  1.44    &    0.20          & -2.62  &    0.15         \\      
                                            &    $R$        &   57559.3413   &  -79.66            &  1.07    &     0.13         & -3.46  &  0.10           \\      
                                            &    $V$        &   57559.4309   &  -25.40            &  2.44    &     0.14         & 2.28  &    0.11         \\      
                                            &    $B$        &   57559.4329   & -23.19             &  1.14    &       0.19       & 2.47  &    0.14         \\      
                                            &    $R$        &   57559.4348   & -20.90             &  2.88    &    0.12          &  2.65 &    0.09         \\      
                                           &     $V$       &  57562.2912     &    -91.09             &   -0.91   &       0.98       & -4.07  &   0.53          \\     
                                            &     $B$       &  57562.2931     &  -90.62               &  -0.81    &        0.97      &  -2.70 &    0.91         \\     
                                            &     $R$       &  57562.2950    &   -90.15              &   0.17   &     0.46         &  -3.54 &   0.36          \\     
                                            &     $V$       &  57562.3048     &   -87.68              & 0.36     &   0.28           & -3.20  &  0.25           \\     
                                            &     $B$       &  57562.3068     &    -87.18             &   1.08   &    0.47          & -2.39  &  0.34           \\     
                                            &     $R$       & 57562.3087     &    -86.67             &  0.09    &    0.27          & -3.90  &    0.18         \\     
                                            &     $V$       &  57562.4243    &      -23.60           &   2.35   &    0.21          & 2.56  &   0.16          \\     
                                            &     $B$       &  57562.4262    &     -21.35            &  1.31    &    0.25          & 3.10  &   0.21          \\     
                                            &     $R$       &  57562.4282    &     -19.05            &  2.44    &     0.17         & 2.87  &   0.13          \\     
 \hline
\end{tabular}
\normalsize
\end{table*}

\begin{table*}
 \centering
  \caption{ Full log of polarised standard star observations used in this document. $q,u$ values are the measured instrumental values in the EFOSC2 
  coordinate system.  ``mid'' denotes the value at the middle of the polarimetric sequence of four exposures.
 \label{table:pol}}
 \begin{tabular}{llllllllll} 
  \hline
Object                                   & Filter   &   Obs. date      & Parallactic angle & $q$                        & $q$ error                & $u$                        & $u$ error \\ 
                                               &            &   (mid, MJD)    & (mid; degrees)    &  ($\times 100\%$)   &  ($\times 100\%$)   & ($\times 100\%$)    & ($\times 100\%$)   \\
 \hline  
BD$-$12 5133                      &    $V$      &  57562.2671     &    139.50            &  7.00    &     0.10         & 2.86  &  0.07           \\
                                             &    $V$      &  57562.2688     &    138.54             &  7.06    &    0.10          & 2.89  &    0.09         \\
                                             &    $B$      &  57562.2707      &   137.59              &  6.23    &    0.19          & 3.80  &   0.14          \\
                                             &    $B$      &  57562.2724      &   136.71              &  6.31    &   0.34           & 3.52  &    0.28         \\
                                             &    $R$      &  57562.2743      &   135.82              &   7.54   &     0.17         & 1.19  &   0.14          \\
                                             &    $R$      &  57562.2760      &   135.01             &  7.66    &   0.13           &  1.25 &   0.07          \\
Hilt 652                                 &    $V$      &   57562.2486      &   97.52              &   0.31   &   0.13           & 3.67  &   0.09          \\
                                             &    $V$      &   57562.2504      &   97.64              &   0.27   &    0.07          & 3.75  &    0.06         \\
                                             &    $B$      &    57562.2522     &   97.77              &  -0.56    &    0.21          & 3.67  &      0.17       \\
                                             &    $B$      &    57562.2540     &   97.89              &  -0.62    &    0.14          & 3.44  &   0.11          \\
                                             &    $R$      &    57562.2559     &  98.03               &  0.92    &   0.09           & 2.87  &    0.07         \\
                                             &    $R$      &    57562.2577     &  98.16               &  0.89    &    0.05          &  2.78 &   0.04          \\
Vela 1 95                              &    $V$   &   57558.9728     &    84.17             &   1.45   &    0.12          & 4.65  &   0.09          \\
                                             &      $B$    &   57558.9747    &     84.71             &   0.19   &    0.36          & 4.17  &   0.27          \\ 
                                             &      $R$    &   57558.9765    &     85.24             &  1.64    &    0.06          & 3.47  &    0.05         \\
                                             &      $V$     &   57559.9767    &    86.07              &  1.39    &    0.12          & 4.68  &    0.09         \\
                                             &      $B$     &   57559.9785   &    86.59              &  0.40    &    0.33          & 4.85  &      0.26       \\
                                             &      $R$     &   57559.9804    &   87.10              &   1.76   &    0.06          & 3.39  &   0.05          \\    
                                             &      $V$     &  57561.9775    &    87.80              &   1.55   &     0.12         & 4.55  &    0.09         \\
                                             &      $B$     &   57561.9793   &     88.32             &  0.81    &    0.37          & 5.35  &     0.29        \\
                                             &      $R$     &   57561.9812   &     88.82            &   2.09   &     0.06         & 3.33  &    0.05         \\      
 \hline
\end{tabular}
\normalsize
\end{table*}

\section{Calibrator selection}
\subsection{Unpolarised standards}\label{sec:unpolarisedstds}
We select a set of unpolarised standard stars (all white dwarfs, see Table \ref{table:obslog1}), which satisfy the following constraints: $V\geq11$ mag (e.g. to avoid non-linearity problems at
lower than average seeing), 
airmass $\leq1.8$, covering a range of parallactic angles at our observing nights, and which have recent observational confirmation of their status as polarimetric calibrator. The selected sources 
are shown in Table \ref{table:obslog1}. All sources, with exception of WD\,1344$+$106, are confirmed zero-polarisation calibrators in the thorough study of \cite{Fossati}. WD~1344$+$106    
has been studied by \cite{Zejmo}, who confirm its suitability as zero polarisation calibrator. Presence of bright Moon, cloud cover and 
the general paucity of faint southern-hemisphere calibrators limited the choice of sources, and the range of parallactic angles at which we could obtain measurements.

\subsection{Polarised standards}\label{sec:polarisedstds}
The selection criteria of intrinsically polarised standards  are similar to the unpolarised ones, but here we point out that the number of (relatively) faint, well-studied, polarised standard stars in the Southern
hemisphere, visible at the time of our observations, is very small. This problem was enhanced by the unlucky coincidence that most of the few suitable sources were too close to the Moon on the first two
nights, and planned observations on the last two nights were hindered by clouds. Therefore only a small sample of polarised standards could be observed (Table \ref{table:obslog1}),
with a narrow range of parallactic angle (Table \ref{table:pol}). All of these objects are listed in \cite{Fossati} with $B,V,R$ band values for $q,u$, from both FORS imaging polarimetry and
spectropolarimetry convolved over synthetic bandpasses; 
note that \cite{Fossati} show some evidence that Vela 1 95 may exhibit low level polarimetric variability.  For completeness, we also list the $q,u$ values of \cite{Fossati} for our standard stars in Table \ref{table:obslog1}. 
For a first order calibration of the instrument response for a Nasmyth-mounted polarimeter it is not strictly required to observe polarised standards, but they do form a valuable additional cross-check.

\section{Observations}\label{sec:obs}
All observations were obtained on the nights of 19, 20 and 22 June 2016. Observations on the nights of 23 and 24 June were impossible
due to poor weather conditions; observations in the second half of 22 June were taken through thick and variable  clouds. The nights of 
19 and 20 June had some thin hazy clouds, and poor seeing conditions.  In all cases, observations used broadband $B$, $V$ and $R$ filters (ESO filters \#639, 641, 642, respectively). 
We obtained no $U$ band standard star data because of the full Moon and cloud conditions, and no $i$ band data 
because fringing starts to become a complication at red wavelengths.

For all our data, we use four half wave plate angles, of 0, 22.5, 45 and 67.5 degrees, taken consecutively. 
The use of four angles increases polarimetric accuracy, as outlined in detail in \cite{patat}. 
The CCD was read out using $2\times2$ binning, the readout mode was  the ``normal'' mode 
(see EFOSC2 user manual, \citealt{usermanual}). We used 
Janesick's method (Janesick 2001) to verify the gain and readnoise in this readout mode and binning, which gave a gain of 1.18 electrons per ADU 
and readnoise of 11 electrons. 

The acquisition of the objects was done as follows: a short exposure without mask and Wollaston element was taken, in which the target was identified, and, after centroiding, placed on
a reference pixel position. This position was determined in the daytime, using an exposure of an internal instrument lamp, illuminating the mask. This allowed us to choose
a position near the center of the CCD but sufficiently away from bad columns and bad quality pixels. We used the pixel position (in unbinned, $1\times1$,  image pixel coordinates) 
of $x,y=1100,1016$. 

To reduce the data, we employ a set of bias frames and a set of polarimetric flat field images. The latter were dome screen flats, taken with the polarimetry optics in the beam 
(Wollaston, wave plate, mask) and the same readout parameters as the science data.  During these flatfield exposures, the wave plate was kept rotating in a continuous fashion. This largely 
scrambles the polarisation-dependent sensitivity properties in the flat fields. The science data were bias-subtracted and flat fielded using standard routines in IRAF. 

The fluxes of the sources in the $o$ and $e$ beam were measured using aperture photometry. The known offset between the $o$ and $e$ beam was used to fit the centroid of the stars more accurately. We used  
our IRAF package {\sc appola} (originally developed by E.~Rol), the method is described in detail in \cite{Wiersema121024} and \cite{Wiersema091018}. 
Aperture radii were chosen as 1.5 times the FWHM of the stellar point spread function, determined with a Gaussian fit. Note that no difference between
$o$ and $e$ beam FWHM distribution was observed at the standard star location, and so we use FWHM values that are found from a tied fit for the 
two beams. To obtain Stokes  parameters $q = Q/I$ and $u = U/I$, and reliable uncertainties, from these fluxes and flux uncertainties, we use the methods in \cite{patat}, also summarised
in \cite{Wiersema091018}. The resulting $q,u$ values are tabulated in Tables 2 and 3, and shown in Figure \ref{fig:matrixmethod}. 

There are some additional complications that affect 
a small number of individual datasets or images.
 Some of the data shown here were taken under very poor or highly variable cloud cover, leading to (relatively) large uncertainties in the $q,u$ values (Table 1 and 2). Several
of the standard stars are very bright, and when observed under poor seeing the relatively small strip height means that
some tweaking of the sky annulus was occasionally required, to prevent the sky annulus from incorporating some pixels from the mask gaps.  Lastly, a small fraction of the data of very 
bright objects occasionally show the presence of ghost images (see EFOSC2 user manual; \citealt{usermanual}), that move as a function of wave plate rotation angle. We carefully analysed each image by hand (and inspected visually) to minimise the effects on the presented analysis.

\subsection{Off-axis behaviour \label{sec:off-axis}}
The observations above were aimed at calibrating the instrumental polarisation near the centre of the CCD - at the reference pixel position, to be precise, where we located
all our science targets too (Higgins et al. in prep.). Many polarimeters show increasing instrumental polarisation away from the optical axis, usually the shape and amplitude of this pattern is a function 
of wavelength, see e.g. \cite{patat} for an example of the 
FORS1 polarisation pattern, or \cite{Heidt} for the pattern of the CAFOS instrument on the Calar Alto 2.2m. To calibrate this, one can use a variety of methods: {\em i)} the sky background in regions with few objects and away from bright Moon; {\em ii)} bright fieldstars in long-exposure science datasets; and {\em iii)} a dedicated observing block of, for example, an open cluster, which has several bright member stars spread over
the field of view, that by virtue of membership of this cluster suffer from broadly identical Galactic dust induced polarisation. In this document we focus our efforts 
on point sources near the center of the field of view, we will investigate the off-axis behaviour in a future study.

\section{Analysis}
\subsection{Fitting a sinusoidal relation \label{sec:cosine}}
As discussed by \cite{Heidt}, the EFOSC2 instrumental polarisation appears to follow a simple cosine curve as a function of 
parallactic angle, with little evidence for additional components of instrumental polarisation. 
 We fit the $q,u$ of the unpolarised standards as function of parallactic angle, using a cosine function, fit independently to $q$ and $u$ to test consistency,
i.e. the function $A*\cos(2\theta - \theta_0)$. For $B$ band, this gives for $q$: $A=2.86\pm0.05, \theta_0=104.3\pm1.2$; for $u$: $A=3.07\pm0.04, \theta_0=13.5\pm0.65$.
For $R$ band, this gives for $q$: $A=3.80\pm0.05, \theta_0=85.9\pm0.7$; for $u$: $A=3.73\pm0.04, \theta_0=-2.6\pm0.6$.
For $V$ band, this gives for $q$: $A=-3.27\pm0.06, \theta_0=274.3\pm1.3$; for $u$: $A=3.32\pm0.04, \theta_0=4.5\pm0.6$.  
 It is clear that the $B,V,R$ filters show polarisation curves with different amplitudes and 
with a phase difference, as 
expected from the properties of metallic mirrors, and as seen in other Nasmyth polarimeters (e.g. \citealt{PAOLO} and references therein). 
For a given wavelength, we expect the cosine amplitude to be the same for $q$ and $u$, and for the $q$ and $u$ data to lag each other by exactly $\pi/2$ phase. We can 
therefore fit $q$ and $u$ together per band, fitting only for the amplitude and one $\theta_0$ value (with a fixed $\pi/2$ phase difference between $q$ and $u$). For $B$ band this gives $A=2.97\pm0.28, \theta_0=103.8\pm0.5$. For $R$ band this gives $A=3.30\pm0.24, \theta_0=94.4\pm0.5$. For $V$ band this gives $A=3.75\pm0.21, \theta_0=86.9\pm0.4$.

The best-fitting cosine function $q,u$ values at a given parallactic angle can now be subtracted off measured $q,u$ values of science objects to
provide a crude correction for instrumental polarisation (as done in e.g. \citealt{Heidt}). However, this would not take into full account the effects of 
cross-talk, which will be substantial (e.g. \citealt{Tinbergen}), particularly affecting intrinsically highly polarised sources. 
A more thorough modelling of this dataset can solve for this too, and give increased accuracy, as we set out in the next section.

\subsection{A Mueller matrix approach}\label{sec:mueller}
Fitting an empirical function as in Section \ref{sec:cosine} is a data-intensive effort, requiring many data points, often impractical for
programmes with only small time allocations. Additionally, it would leave smaller instrumental effects (cross-talk) incompletely corrected. 
An alternative approach that is frequently used for Nasmyth focus mounted polarimeters,  is one where all optical elements in the light path are expressed as Mueller matrices, acting  upon the Stokes vector describing the incoming light (see e.g. \citealt{Tinbergen} and 
references therein). Naturally, the tertiary mirror (M3) is the most important contributor, as these are metal-coated mirrors under a large 
angle, and therefore strongly polarise incoming light. We construct a train of matrices describing all key polarising components of 
EFOSC2, and fit the unknown quantities (e.g. the complex index of refraction $n_c$ [defined in terms of the refractive index $n$ and the extinction coefficient $k$  as $n_c = n - i*k$], the possible angular offsets between the 
detector and the celestial reference, etc.) onto the dataset described above. The coating of M3 slowly 
oxidises and dust  is accumulated on the mirror surface, and we therefore expect these indices to gradually change with time and potentially change abruptly any time the mirror is recoated or washed
(e.g. \citealt{vanHarten}). The same is 
true any time the instrument is subject to an important maintenance operation that can modify the angular offset of the optical components of the whole 
instrument. The calibration of a Nasmyth polarimeter for relatively simple alt-azimuthal telescopes and instruments by means of suitable trains of Mueller 
matrices was addressed by several authors \citep{Giro,Joos,PAOLO} while more complex instruments were also successfully modelled \citep{Selbing,Witzel}. 

\begin{figure*}
\centerline{\includegraphics[width=19.5cm]{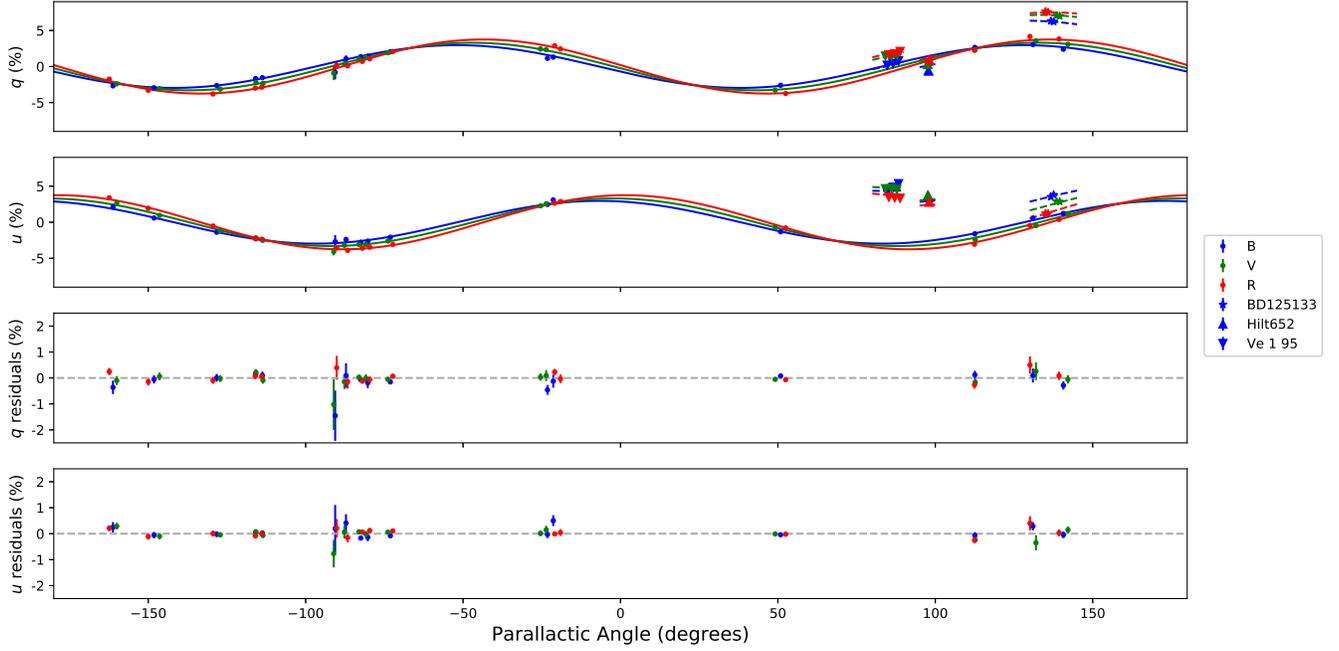}}
\caption{ The measured $q,u$ values in $B,V,R$ bands of unpolarised standards (circles) and polarised standards (triangles, stars) are shown as a function of parallactic angle (PA) in the top two windows. The solid lines show the best fitting $B,V,R$ polarimetric Mueller matrix model solutions; the dashed lines show the same solution around the polarised stars values (using shorter lines to keep the plot legible). The bottom two windows show the residuals for $q,u$ in $B,V,R$ bands. The average $V$ band residuals of the $q$ and $u$ fits are calculated to be $\sim0.06\%$, with similar values for $B,R$ bands.}
\label{fig:matrixmethod}
\end{figure*}

\begin{figure}
\includegraphics[width=\columnwidth]{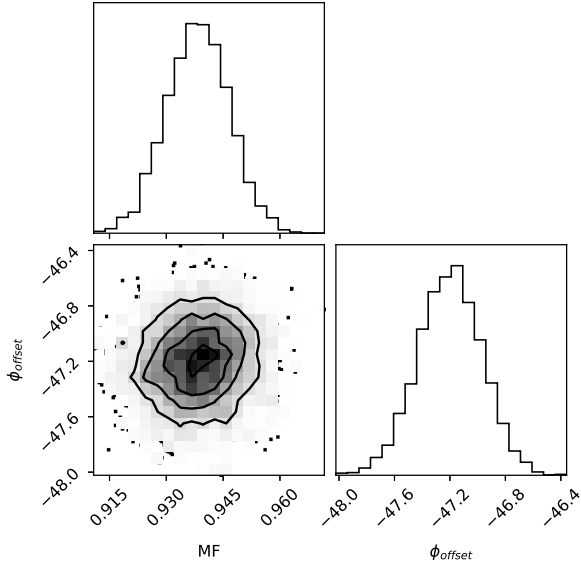}
\caption{ Projection of the normalised probability distributions for the detector offset angle $\phi_{\rm offset}$ and multiplication factor (MF), from the MCMC analysis, for the $V$ band dataset.}
\label{fig:mcmcresult}
\end{figure}

For our calibration effort of EFOSC2 we make use of a Mueller matrix train consisting of the following component matrices, closely following the setup for the 
Telescopio Nazionale Galileo (TNG) described in \cite{Giro}:
\begin{itemize}
\item{Matrix representing the incoming source light from the M2 mirror onto the M3 mirror: $[M_{M3}(0^{\circ})]$}. The purpose of this matrix is to change the sign of $U$.
\item{Rotation matrix representing the transformation from the sky coordinates reference frame into mirror coordinates frame as a function of telescope pointing direction: $[T(-\theta_{\rm pa})]$}
\item{Matrix representing the physical properties of the M3 mirror (including the value of, and dependence on wavelength of, $n$ and $k$, mentioned above) and the $45^{\circ}$ reflection due to how the mirror is mounted: $[M_{M3}(45^{\circ})]$}. 
\item{Rotation matrix representing the transformation for a change in elevation of the mirror as a change in the mirror orientation with respect to the derotator focal plane: $[T(-\theta_{\rm pa})]$}
\item{Rotation matrix representing the transformation from the mirror reference frame to the reference frame of the detector: $[T(\phi_{\rm offset})]$}.
\end{itemize}
Here $\theta_{\rm pa}$ is the parallactic angle. 
This leads to a Mueller matrix for the telescope represented by the following equation:
\begin{align*}
[M_T] &= [T(\phi_{\rm offset})] \times [T(-\theta_{\rm pa})] \times [M_{M3}(45^{\circ})] \\
&\times [T(-\theta_{\rm pa})] \times [M_{M3}(0^{\circ})]
\end{align*}
where $[M_T]$ is the total matrix representing the telescope and EFOSC2,  and the contributing matrices are described above.

We use the prescription of \cite{Stenflo} to numerically evaluate the matrix components describing the M3 mirror, and use the material constants as a function of wavelength for pure aluminium as 
tabulated by \cite{Rakic}, to describe the aluminium coating of M3.  A single numerical multiplication factor (an efficiency factor, as it were) is used, which adjusts the effective refractive index of the aluminium mirror to account for the fact that the mirror surface is not ideal, pure, aluminium, but shows effects of oxidation and dust (and perhaps other effects that can be caught with this simple parametrisation). This simple approach is also successfully used for e.g. the PAOLO instrument (\citealt{PAOLO}).

Using the M3 physical components described above, we give as illustration the following matrices for a reflection at $0^{\circ}$
\[
[M_{M3}(0^{\circ})] = 
\left[ \begin{array}{cccc} 1 & 0 & 0 & 0 \\ 0 & 1 & 0 & 0 \\ 0 & 0 & -1 & 0 \\ 0 & 0 & 0 & -1 \end{array} \right]
\]
and a reflection at $45^{\circ}$
\[
[M_{M3}(45^{\circ})] = 
\left[ \begin{array}{cccc} 0.9699 & 0.0301 & 0 & 0 \\ 0.0301 & 0.9699 & 0 & 0 \\ 0 & 0 & -0.9487 & -0.1993 \\ 0 & 0 & 0.1993 & -0.9487 \end{array} \right]
\]
for $V$ band - these values are different for $B$ and $R$ bands as the material constants are wavelength dependent.  

The equations for the matrix components were implemented in a Python code, in which we fit the values of the matrices onto the large and high quality set of polarised and unpolarised standard stars described above.
This allowed us to derive a very well constrained polarimetric model (Figure \ref{fig:matrixmethod}), using only a small number of free parameters. It simultaneously corrects for the instrumental polarisation and the angular offset of the detector through the entire possible range 
of the parallactic angle of any observable target. The root mean square (rms) of the residuals of the best fitting model on the observed data (Figure \ref{fig:matrixmethod}) are consistent with being due to the observational errors only. The set of polarised standard stars similarly shows excellent agreement with our model (Figure \ref{fig:matrixmethod}). 
To get a better understanding of any possible degeneracies and shape of confidence contours,  we run a Markov Chain Monte Carlo (MCMC) code ({\tt emcee}; \citealt{emcee}).  We use the log-likelihood function of a normal distribution as our posterior probability distribution and attempt to find the maximum-likelihood result. We use this in conjunction with the observed $q$ and $u$ values to determine the detector offset angle with respect to the mirror's reference frame (wavelength dependent) and the afore mentioned multiplication factor.  The MCMC uses 20 walkers, each with a total of 2500 steps and a burn-in period of 250 steps. Figure \ref{fig:mcmcresult} shows the projection of the probability distributions of both the offset angle and multiplication factor (MF) for $V$ band. The parameters are clearly non degenerate, and both distributions are consistent with normal, and show low levels of variance. We see similar distributions for the parameters for both the $B$ and $R$ bands and are therefore confident that our results accurately reflect the true offset angle and multiplication factor. The resulting full calibration parameter values from the MCMC analysis can be seen in Table \ref{table:mcmc_results}.

\begin{table}
\centering
\caption{Detector angle offset and multiplication factor results from the MCMC code, for the $B$, $R$ and $V$ filters. Errors quoted are $1\sigma$.}
\begin{tabular}{ccc}
    \hline
  \multicolumn{1}{p{2.0cm}}{\centering Filter}
& \multicolumn{1}{p{2.0cm}}{\centering $\phi_{\rm offset}$ ($^{\circ}$)} 
& \multicolumn{1}{p{2.0cm}}{\centering Multiplication \\ factor}\\ \hline
$B$ & $-51.9\pm0.3$ & $0.95\pm0.01$ \\ 
$V$ & $-47.2\pm0.3$ & $0.94\pm0.01$ \\ 
$R$ & $-43.5\,^{+0.2}_{-0.3}$ & $0.94\pm0.01$ \\ \hline
\end{tabular}
\label{table:mcmc_results}
\end{table}

To calibrate $q,u$ observations taken with EFOSC2, one can now invert the matrix model found above to directly get the instrument-corrected $q,u$ values from the 
observed values, using the parallactic angle value. 
This model could in principle be extended to spectropolarimetry or to other wavelengths (filters) with small modifications or additions to the code.

\subsection{Comparison with older observations}
\begin{figure}
\begin{center}
\includegraphics[width=7cm]{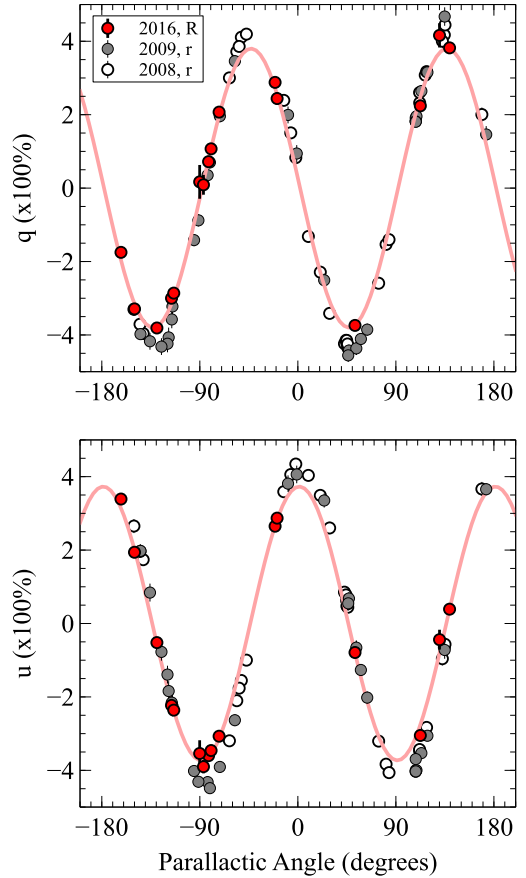}
\caption{Shown in red are the zero-polarisation standard star datapoints, in instrumental coordinates, from our programme; 
in grey and white are the data points from Heidt \& Nilsson (2011). The thin red line is a cosine fit to our data, to make it easier to see the difference between the data from 2016 and those from 2008 and 2009. A  significant change in amplitude can easily be seen, a small phase shift may also be present. }
\label{fig:Rcomp}
\end{center}
\end{figure}

An instrumental polarimetry characterisation of EFOSC2, mounted on the NTT, was undertaken by \cite{Heidt} using data taken in
2008 and 2009. These authors obtained a large number of observations of zero-polarisation standards in a single broadband filter, and find that the
observed instrumental $Q,U$ values as a function of parallactic angle can be well described by a cosine function. 
In principle the data from \cite{Heidt} provide a meaningful comparison with our data, but there are some complications:
 \cite{Heidt} used a Gunn $r$ filter (ESO filter \#786, decommissioned in 2009); and there were
 two rounds of NTT mirror re-coating in between the Heidt \& Nilsson observing dates and ours, on 3-7 July 2012 and 
 4-12 May 2015 (M1 + M3 mirrors). In Figure \ref{fig:Rcomp} we plot our zero-polarisation standard star $R$ band data, together with the $r$ standard star observations from \cite{Heidt}, which were taken on two epochs. As is clearly
 visible, the amplitude of instrumental polarisation appears to change significantly between epochs, a small change in phase may also be present. This demonstrates that care needs to be taken when using older archival data to calibrate new data. 

\section{Discussion and conclusions}\label{sec:discussion}
Our analysis above shows that a calibration of EFOSC2 linear imaging polarimetry (in $B,V,R$), using a relatively simple Mueller matrix based instrument model, is possible to 
a precision of $P \sim0.1\%$:  propagating the uncertainties derived from the MCMC simulations shown in Figure \ref{fig:mcmcresult} and discussed in Section \ref{sec:mueller}, gives a value of the calibration accuracy in $V$ band of $P=0.08\%$. 
This level is sufficient for the majority of the most popular science applications of EFOSC2 imaging polarimetry. 

There are some important caveats and recommendations to point out here:
\begin{unnumlist}
\item While the model we built works very well with our measurements, we note that we have no measurements
of Stokes $v$ $(=V/I)$; the quarter wave plate was not mounted for our observing programme, which concerned linear polarimetry only. 
The M3 reflection should produce large amounts of circular polarisation as a function of parallactic angle and wavelength, as seen by e.g. PAOLO on the TNG (\citealt{PAOLO}).
Measurements of $v$, preferably quasi-simultaneously (within a few days) with $q$ and $u$, would provide a strong additional calibration constraint,
testing the accuracy of the crosstalk terms further.
\item In their analysis of the instrumental polarisation of the PAOLO instrument on the TNG telescope, \cite{PAOLO} note that 
the typical timescales through which a given calibration is valid is days to weeks. In that sense it is not surprising to find  small but significantly different instrumental polarisation values from \cite{Heidt}. The timescales of change of the instrumental polarisation
could be monitored through infrequent calibration plan observations. Using our calibration as a lead, this would require only a small number of standard star observations, and would allow more reliable re-processing of past datasets. 
\item The phases of the instrumental $q,u$ curves (Figure \ref{fig:matrixmethod}) are surprisingly strongly dependent on wavelength,
more so than seen in e.g. PAOLO (\citealt{PAOLO}). We only cover the $B,V,R$ filters here, future observations covering also $U$ and $i$ filters would establish
this behaviour over a wider wavelength range.
\item As can be seen in e.g. Fig \ref{fig:matrixmethod}, our observations of polarised standards are limited to a small range of 
parallactic angle, and we only observed three sources. The underlying cause of this is that the number of polarised standard stars 
than can be observed from the Southern hemisphere is very small, and many are too bright for a 4m class telescope. During our observing
nights the choice was therefore limited. 
\item Some improvements in accuracy of individual standard star $q,u$ values are possible. Putting aside the obvious advantages of
observations in clear and lower moonlight conditions, an increase in accuracy can also be delivered by using eight wave plate angles 
instead of the four we used. Using a small amount of defocussing of the telescope will allow longer exposures and will spread the light over
more pixels, allowing more counts in the source PSF without saturating the central pixel. This latter effect can be important on nights
of good seeing, considering the brightness of the standards.
\item An explicit matrix for the half wave plate could be added. This could reduce the uncertainties that come from the fact that a given set of four consecutive wave plate angle exposures, needed for a single $Q,U$ point, spans a (small) range of parallactic angle.  
\end{unnumlist}

While we only considered broadband imaging polarimetry in $B, V, R$ filters, a similar strategy for narrowband and spectropolarimetry can be employed with relatively minor adjustments to our Python codes.

\begin{acknowledgements}
This calibration effort is dedicated to the memory of dr.~Peter Curran (ICRAR, Perth). We miss his insights into polarisation properties of jet sources, but above all his generous friendship. We thank the anonymous referee for their helpful report which improved this paper. 
It is a pleasure to thank the observatory staff at ESO La Silla for their help in obtaining the observations in this
document, and particularly Pablo Arias for a tour of the telescope and detailed explanation of the instrument. We are very grateful to J. Heidt and K. Nilsson for 
sharing their standard star measurements with us. 
Based on observations collected at the European Organisation for Astronomical Research in the Southern Hemisphere under ESO programme 097.D-0891(A).
KW, ABH and RLCS acknowledge funding from STFC. 

\end{acknowledgements}

\bibliographystyle{pasa-mnras}

\end{document}